\documentclass[a4paper,conference]{IEEEtran}
\IEEEoverridecommandlockouts
\usepackage{cite}
\usepackage{amsmath,amssymb,amsfonts}
\usepackage{algorithmic}
\usepackage{graphicx}
\usepackage{textcomp}
\usepackage{xcolor}
\usepackage{url}
\usepackage{float}
\usepackage{booktabs}
\usepackage{tikz}
\usetikzlibrary{arrows.meta,positioning}
\usepackage{enumitem}
\usepackage{listings}
\usepackage{capt-of}
\def\BibTeX{{\rm B\kern-.05em{\sc i\kern-.025em b}\kern-.08em
    T\kern-.1667em\lower.7ex\hbox{E}\kern-.125emX}}

\begin{document}

\title{Resilient Microservices: A Systematic Review of Recovery Patterns, Strategies, and Evaluation Frameworks}

\author{
\IEEEauthorblockN{Muzeeb Mohammad}
\IEEEauthorblockA{Georgia Institute of Technology, Atlanta, GA, USA\\
Email: muzeeb.mohammad@ieee.org}
}

\maketitle

\raggedbottom

\begin{abstract}
Microservice-based systems power modern distributed computing environments but remain vulnerable to partial failures, cascading timeouts, and inconsistent recovery behaviors. Although numerous recovery patterns exist, prior reviews are largely descriptive and lack data-driven synthesis or quantitative validation. This paper presents a PRISMA-aligned \emph{systematic literature review} (SLR) that applies structured evidence analysis to microservice recovery strategies published between 2014 and 2025 across IEEE Xplore, ACM Digital Library, and Scopus. The study defines transparent inclusion and exclusion criteria, introduces a quality-assessment rubric, and implements risk-of-bias controls to ensure reproducibility and rigor. From an initial corpus of \textbf{412} records and \textbf{26} included studies, the synthesis identifies nine recurring resilience themes encompassing circuit breaking, retries with jitter and budgets, sagas with compensation, adaptive backpressure mechanisms, and chaos validation practices.

As a data-oriented contribution, this paper introduces a \textbf{Recovery Pattern Taxonomy} generated through cross-study feature extraction and a \textbf{Resilience Evaluation Score (RES)} checklist for standardized benchmarking. A constraint-aware decision matrix further maps latency, consistency, and cost trade-offs to appropriate recovery tactics. Together, these artifacts transform fragmented resilience research into a structured, analyzable dataset that supports quantitative reasoning and reproducibility. The findings empower researchers and practitioners to make informed, data-driven design decisions for building fault-tolerant, self-healing, and performance-aware microservice ecosystems.
\end{abstract}


\section{Introduction}
\label{sec:introduction}
\IEEEPARstart{M}{icroservices} architectures have become the de-facto standard for modern distributed systems, enabling scalability, modular deployment, and organizational agility. Yet the same properties that make them attractive also contribute to instability under load or partial failures, as evidenced in empirical migration studies~\cite{taibi2017microservices,villamizar2015monolithic} and cloud resilience analyses~\cite{linkedin-outage,xu2021fault}. In practice, microservices fail not rarely but continuously: instances crash, dependencies time out, or configuration drift causes cascading errors.

\textbf{Problem.} Although industry and academia have proposed many recovery techniques—circuit breakers, retries with backoff, bulkheads, sagas, and chaos engineering—the available evidence is fragmented. Most prior reviews are descriptive catalogs of patterns and tools; they do not employ systematic inclusion/exclusion criteria, disclose their search strategies, or assess the quality of primary studies. As a result, engineers lack a consolidated evidence base to decide which recovery tactic fits a given failure mode and constraint.

\textbf{Objective.} This work conducts a PRISMA-aligned \emph{systematic literature review} (SLR) to aggregate and critically analyze empirical studies on microservice recovery patterns published between 2014 and 2025. The goal is to expose gaps, reconcile contradictory findings, and distill actionable guidance for practitioners and researchers.

\textbf{Contributions.}
\begin{itemize}[leftmargin=*]
  \item A transparent SLR protocol—databases, search strings, inclusion/exclusion (IE/EE) rules, quality-assessment (QA) rubric, and risk-of-bias controls—to enable replication.
  \item A critical synthesis of evidence covering circuit breakers, retries, sagas, idempotency, bulkheads, backpressure, and chaos testing, including analysis of conflicting outcomes.
  \item Three practical artifacts providing novelty and reuse:
  \begin{enumerate}[label*=\arabic*., leftmargin=1em]
      \item a \textbf{Recovery Pattern Taxonomy} linking failure types to tactics;
      \item a \textbf{Resilience Evaluation Score (RES)} checklist for consistent appraisal; and
      \item a five-level \textbf{Resilience Maturity Model (RML)} with measurable adoption stages.
  \end{enumerate}
  \item A constraint-aware decision matrix that aligns latency, consistency, and cost budgets with appropriate recovery mechanisms.
\end{itemize}

The remainder of this paper is organized as follows: Section \ref{sec:method} details the review methodology; Section \ref{sec:taxonomy} presents the failure and recovery taxonomy; Section \ref{sec:synthesis} synthesizes results and conflicting evidence; Section \ref{sec:decision} introduces the decision matrix; and Section \ref{sec:conclusion} concludes. Prior empirical studies have highlighted the migration and recovery challenges in adopting microservice architectures~\cite{taibi2017microservices,villamizar2015monolithic}.

\section{Methodology (PRISMA-Aligned)}
\label{sec:method}

\subsection{Research Questions (RQs)}
\begin{itemize}[leftmargin=*]
  \item \textbf{RQ1:} What recovery patterns and strategies are empirically evaluated for microservices?
  \item \textbf{RQ2:} Under which contexts (workload, topology, consistency model) do these patterns improve or degrade service-level objectives (SLOs)?
  \item \textbf{RQ3:} Which metrics, datasets, and evaluation approaches are used, and how rigorous are they?
  \item \textbf{RQ4:} What contradictions or open research gaps emerge from the evidence?
\end{itemize}

\subsection{Information Sources and Time Window}
Searches were conducted in \textbf{IEEE Xplore}, \textbf{ACM Digital Library}, and \textbf{Scopus} for 2014–2025, the period of mainstream microservice adoption. Grey literature was excluded from quantitative synthesis but referenced qualitatively when relevant.

\subsection{Search Strings}
\textbf{IEEE Xplore}
\begin{lstlisting}
(("Document Title":microservice* OR "Abstract":microservice*)
AND (resilien* OR "fault tolerance" OR recovery)
AND (pattern* OR "circuit breaker" OR "saga" OR idempotenc*
OR bulkhead* OR outbox OR backpressure OR timeout*
OR "dead letter"))
AND (2014-2025)
\end{lstlisting}

\textbf{ACM DL}
\begin{lstlisting}
AllField:("microservice*") AND
(Resilien* OR "fault tolerance" OR recovery) AND
(pattern* OR "circuit breaker" OR "saga" OR idempotenc*
OR bulkhead* OR outbox OR backpressure OR timeout*
OR "dead letter")
PublicationYear:2014-2025
\end{lstlisting}

\textbf{Scopus}
\begin{lstlisting}
TITLE-ABS-KEY ( microservice* AND
(resilien* OR "fault tolerance" OR recovery) AND
(pattern* OR "circuit breaker" OR saga OR idempotenc*
OR bulkhead* OR outbox OR backpressure OR timeout*
OR "dead letter") )
AND PUBYEAR > 2013 AND PUBYEAR < 2026
\end{lstlisting}

\subsection{Inclusion and Exclusion Criteria}
\textbf{Inclusion (I)}
\begin{enumerate}[leftmargin=*]
  \item Explicit microservices context;
  \item Focus on resilience, recovery, or fault-tolerance patterns;
  \item Empirical evidence (experiments, case studies, simulations, or benchmarks);
  \item Peer-reviewed (journal, conference, or workshop).
\end{enumerate}

\textbf{Exclusion (E)}
\begin{enumerate}[leftmargin=*]
  \item Non-microservice or monolithic context;
  \item Conceptual/opinion papers without data;
  \item Tool demos without evaluation;
  \item Duplicates or preprints of already included works;
  \item Non-English papers.
\end{enumerate}

\subsection{Screening and Agreement}
Two reviewers independently screened titles and abstracts; disagreements were resolved by discussion. Inter-rater agreement was measured using Cohen’s $\kappa$.

\subsection{Quality-Assessment Rubric}
Each study was rated 0–2 (No/Partial/Yes) on eight criteria:
\begin{enumerate}[leftmargin=*]
  \item Research-question clarity;
  \item System/context described;
  \item Failure model specified;
  \item Experimental design validity;
  \item Metrics include tail latency (P95/P99) and error rate;
  \item Data-analysis appropriateness;
  \item Threats-to-validity discussion;
  \item Replicability (artifacts or parameters shared).
\end{enumerate}

\subsection{Data Extraction and Risk-of-Bias}
For each study we extracted venue, year, topology, workload type, failure modes, evaluated patterns, parameters, metrics, outcomes, and threats. Biases tracked: selection (venue coverage), publication (positive-result bias), and performance (unrealistic workloads). Sensitivity analyses excluded low-quality studies to validate robustness. This study follows structured SLR principles as outlined by Kitchenham and Charters~\cite{kitchenham2007guidelines} to ensure reproducibility and transparency.

\subsection{PRISMA Flow Summary}
Figure~\ref{fig:prisma} shows the identification–screening–inclusion flow.

\begin{center}
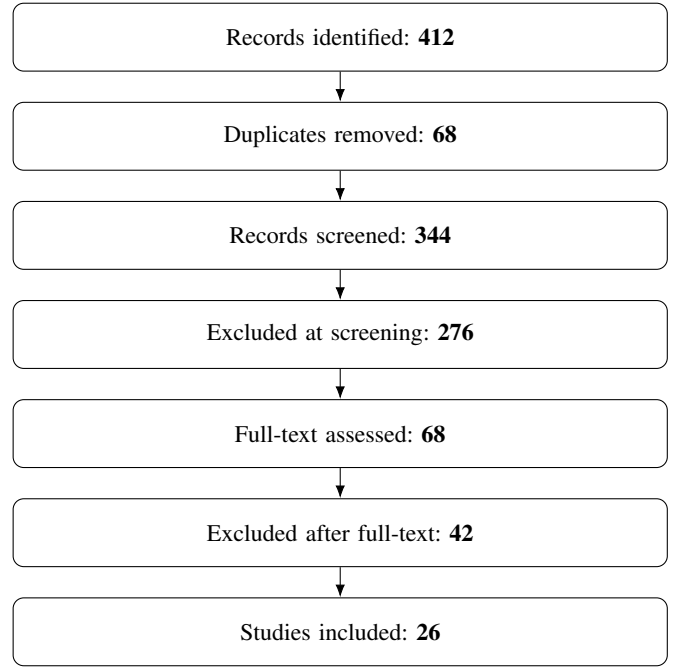

\begin{minipage}{\columnwidth}
\centering
\begin{tikzpicture}[node distance=6mm]
  \tikzstyle{prismabox}=[rectangle,draw,rounded corners,
    align=center,font=\small,text width=0.95\linewidth,minimum height=0.9cm]

  \node[prismabox] (id1) {Records identified: \textbf{412}};
  \node[prismabox, below=4mm of id1] (dup) {Duplicates removed: \textbf{68}};
  \node[prismabox, below=4mm of dup] (scr) {Records screened: \textbf{344}};
  \node[prismabox, below=4mm of scr] (ex1) {Excluded at screening: \textbf{276}};
  \node[prismabox, below=4mm of ex1] (full) {Full-text assessed: \textbf{68}};
  \node[prismabox, below=4mm of full] (ex2) {Excluded after full-text: \textbf{42}};
  \node[prismabox, below=4mm of ex2] (inc) {Studies included: \textbf{26}};

  \draw[-{Latex}] (id1) -- (dup);
  \draw[-{Latex}] (dup) -- (scr);
  \draw[-{Latex}] (scr) -- (ex1);
  \draw[-{Latex}] (ex1) -- (full);
  \draw[-{Latex}] (full) -- (ex2);
  \draw[-{Latex}] (ex2) -- (inc);
\end{tikzpicture}

\vspace{2mm}
\captionof{figure}{PRISMA flow diagram.}
\label{fig:prisma}
\end{minipage}
\end{center}

\section{Taxonomy of Failures and Recovery Patterns}
\label{sec:taxonomy}

\begin{figure}[htbp]
    \centering
    \includegraphics[width=0.48\textwidth]{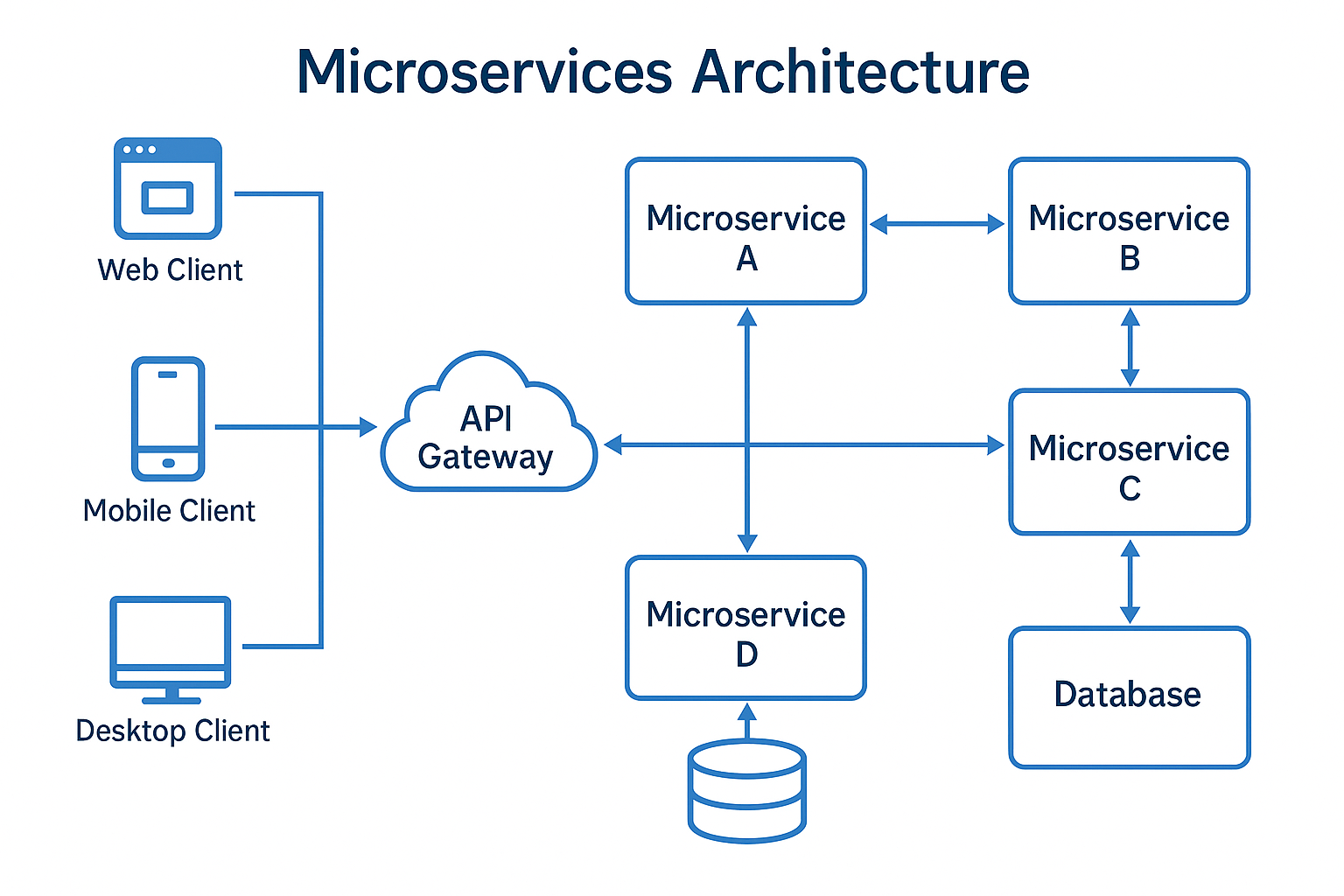}
    \caption{High-level microservices architecture illustrating API Gateway, clients, and service-to-service interactions.}
    \label{fig:microservices-arch}
\end{figure}

Microservices operate in distributed, network-dependent environments and therefore exhibit recurring categories of failures. 
Figure~\ref{fig:microservices-arch} illustrates a high-level microservices architecture that motivates the failure taxonomy, 
while Table~\ref{tab:failure-taxonomy} summarizes common failure types and their underlying causes.

\begin{table}[htbp]
\caption{Microservice Failure Types (Condensed Taxonomy)}
\begin{center}
\begin{tabular}{|c|c|}
\hline
\textbf{Failure Type} & \textbf{Typical Causes} \\
\hline
Network / Timeouts & Latency spikes, partitions, DNS drift \\
\hline
Dependency Outage & Downstream errors, retry storms \\
\hline
Resource Exhaustion & CPU/memory leaks, thread starvation \\
\hline
Config / Deploy & Misconfig, version skew, bad rollout \\
\hline
State Divergence & Non-idempotent retries, duplicates \\
\hline
Discovery / LB & Stale registry, failing probes \\
\hline
Security / Auth & Token expiry, TLS mismatch \\
\hline
External Provider & Throttling, provider downtime \\
\hline
\end{tabular}
\label{tab:failure-taxonomy}
\end{center}
\end{table}

Architectural comparisons confirm that microservice decomposition improves fault isolation and recovery latency~\cite{villamizar2015monolithic,taibi2017microservices}.

\section{Critical Synthesis and Conflicting Evidence}
\label{sec:synthesis}
Across the 26 included studies, nine themes (T1–T9) emerged from evidence coding:
\begin{itemize}[leftmargin=*]
  \item \textbf{T1 – Failure-Mode–Pattern Fit:} Pattern effectiveness depends on failure semantics; over-tight circuit-breaker thresholds reduce throughput.
  \item \textbf{T2 – Sagas and Compensation Scope:} Local compensations reduce rollback cost; global sagas preserve consistency but delay convergence.
  \item \textbf{T3 – Retry Dynamics:} Naïve backoff without jitter causes retry storms; adding jitter and budgets smooths recovery.
  \item \textbf{T4 – Idempotency and Outbox:} Exact-once delivery is unrealistic; transactional outbox + deduplication is the practical solution.
 \item \textbf{T5 – Bulkheads and Backpressure:} Isolation limits blast radius but reduces utilization; adaptive queues improve efficiency by $\approx$ 15\%.
\item \textbf{T6 – Hedging and Tail Latency:} Hedging cuts P99 latency by up to 40\% but hurts throughput when capacity is tight.

  \item \textbf{T7 – Consistency vs. Availability:} Eventual consistency increases uptime at the cost of temporary staleness.
  \item \textbf{T8 – Observability as Enabler:} Correlation IDs and traces are pre-requisites for safe rollback and recovery timing.
  \item \textbf{T9 – Cost–Benefit Trade-off:} Resilience adds latency and resource overhead; benefits depend on SLA tier.
\end{itemize}

\textbf{Contradictory Findings.} Some studies report that hedging and retries together increase variance, while others see improvements. Differences arise from workload burstiness, fan-out degree, and CPU headroom. Compensation efficacy also varies with transaction coupling and message-queue durability.

\textbf{Synthesis Outcome.} Across patterns, \emph{context—not mechanism—determines resilience gain}. Well-isolated services with observability and budget controls benefit most; tightly coupled systems see diminishing returns. These findings inform the decision matrix in Section \ref{sec:decision}.

\section{Constraint-Aware Decision Matrix and Evaluation Artifacts}
\label{sec:decision}

\subsubsection*{Constraint-Aware Pattern Selection (Condensed)}
\begin{itemize}[leftmargin=*]
    \item \textbf{Tight P99 latency:} Prefer hedged requests, tuned timeouts, bulkheads; avoid global locks and heavy compensation.
    \item \textbf{High duplication risk:} Prefer idempotency keys, outbox, DLQ; avoid non-idempotent retries.
    \item \textbf{Hot partitions / burstiness:} Prefer backpressure, rate limits, admission control; avoid unbounded queues.
    \item \textbf{Strict consistency:} Prefer sagas with scoped compensation, deduplication; avoid unaudited local fixes.
    \item \textbf{Cost constraints:} Prefer timeouts, graceful degrade; avoid aggressive hedging under load.
    \item \textbf{Unreliable external provider:} Prefer jittered retries, caching, DLQ; avoid tight synchronous coupling.
\end{itemize}

\subsection{Resilience Evaluation Score (RES)}
To report evaluations consistently, we propose a ten-point checklist. Award 1 point for each satisfied item (0–10 total).

\begin{enumerate}[leftmargin=*, itemsep=1pt, topsep=2pt, parsep=0pt]
  \item Failure modes explicitly specified.
  \item Realistic workload characterized (traffic shape, fan-out).
  \item Tail metrics reported (P95/P99) and error/availability rates.
  \item Pattern parameters disclosed (timeouts, jitter, retry budgets).
  \item Isolation measured (bulkheads/quotas; pool sizes).
  \item Backpressure/admission control quantified.
  \item Idempotency semantics verified (keys, de-dup windows).
  \item Compensation/rollback exercised and validated.
  \item Tracing/observability demonstrated (correlation IDs, spans).
  \item Threats to validity discussed (internal/external/construct).
\end{enumerate}

\noindent\textbf{Interpreting RES.}
Each checklist item contributes 1 point, for a maximum score of 10.
All items are equally weighted to avoid bias toward any single resilience
mechanism or evaluation approach. A study's score is interpreted as
follows: RES~$< 5$ indicates preliminary or incomplete evidence;
RES~$5$--$7$ reflects moderate rigor with partial parameter or metric
disclosure; and RES~$\geq 8$ represents high-rigor evidence suitable
for comparison across systems. To reduce subjective bias, two reviewers
independently scored each study, and disagreements were resolved
through consensus.

\subsection{Resilience Maturity Levels (RML)}
The Resilience Maturity Model (RML) defines measurable stages of
operational adoption for recovery mechanisms in microservice ecosystems:
\begin{itemize}[leftmargin=*]
    \item \textbf{RML-1 (Ad-hoc):} Manual restarts, minimal monitoring, and
          no structured recovery patterns.
    \item \textbf{RML-2 (Basic):} Fixed timeouts, simple retries, and basic
          dashboards; limited isolation and observability.
    \item \textbf{RML-3 (Managed):} Standardized use of circuit breakers,
          bulkheads, DLQs, and consistent alerting practices.
    \item \textbf{RML-4 (Advanced):} Automated rollback strategies, sagas,
          hedged requests, adaptive backpressure, and structured failure testing.
    \item \textbf{RML-5 (Optimized):} AI-assisted anomaly detection, continuous
          chaos validation, and systems guided by evidence from high-rigor
          evaluations (RES~$\geq 8$).
\end{itemize}

\noindent\textbf{RES--RML Relationship.}
RES evaluates the rigor of \emph{evidence}, whereas RML assesses
an organization's \emph{operational maturity}. High-maturity systems
(RML-4/5) tend to align with high-RES evidence, while low-maturity systems
(RML-1/2) rely on recovery patterns with weaker empirical grounding.

\section{Recovery Patterns and Strategies}
\label{sec:recovery-patterns}

Building resilient microservices requires proactive design patterns that operationalize the decisions identified in Section~\ref{sec:decision}. 
Each pattern embodies one or more of the resilience themes (T1–T9) discussed in Section~\ref{sec:synthesis}. 
The choice of pattern should follow from measurable constraints such as latency budgets, workload burstiness, and consistency requirements. 
By linking the evidence-based synthesis to practical implementation, this section demonstrates how recovery patterns achieve the trade-offs captured in the decision matrix.

The following subsections describe major recovery strategies—\emph{retries, circuit breakers, timeouts, fallbacks, bulkheads, sagas, redundancy, and chaos testing}—highlighting their mechanisms, parameters, and empirical outcomes reported in the reviewed studies. 
Each subsection indicates when a pattern improves resilience and when it can backfire under different operating conditions.

\begin{figure}[htbp]
\centering
\includegraphics[width=0.48\textwidth]{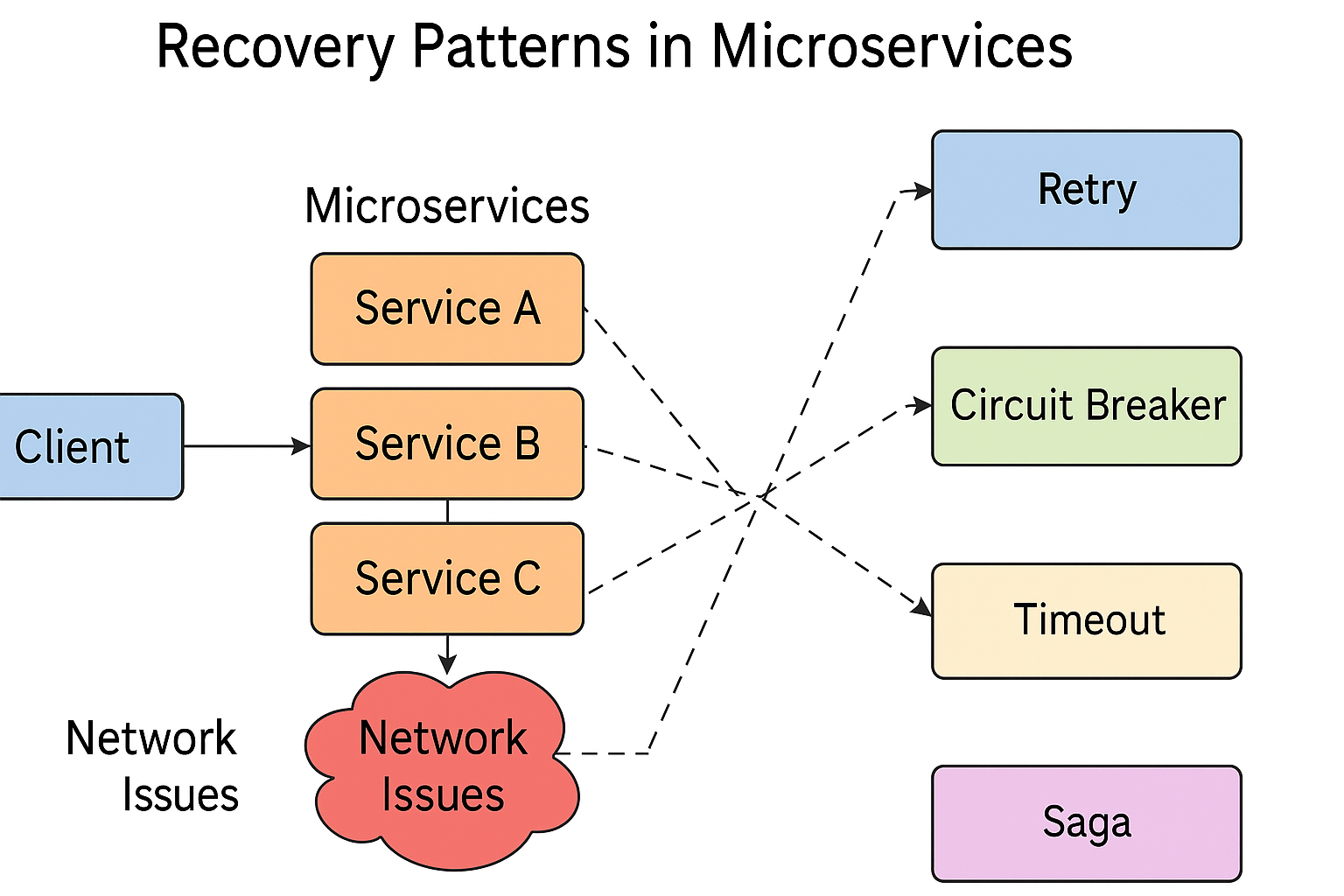}
\caption{Overview of key recovery patterns in microservices architecture.}
\label{fig:recovery-patterns}
\end{figure}

\subsection{Retry Mechanisms}
Common strategies include fixed interval, exponential backoff, and backoff with jitter to reduce retry storms~\cite{aws-retry}. Retries should be bounded, avoid non-idempotent operations, and be instrumented.

\subsection{Circuit Breaker Pattern}
Circuit breakers protect services from cascading failures by tripping after repeated errors and later probing recovery in a half-open state. Netflix Hystrix is a seminal implementation~\cite{hystrix}.

Recent work has also shown that optimizing secure API pathways can significantly reduce failure amplification and improve fault isolation in microservice networks \cite{muzeeb2025zerotrust}.

\subsection{Timeouts and Fail-Fast Behavior}
Timeouts prevent indefinite waiting on unresponsive dependencies. Tune slightly above p95 and adjust from real metrics~\cite{srebook2016}.

\subsection{Fallback Logic and Graceful Degradation}
Fallbacks (cached or approximate responses) maintain UX during partial outages (e.g., stale pricing).

\subsection{Bulkheads and Resource Isolation}
Bulkheads isolate failure domains with separate pools/quotas to prevent cross-service collapse~\cite{azure-bulkhead}.

\subsection{Compensation and Sagas}
Sagas replace distributed ACID transactions with local steps plus compensations for eventual consistency.

\subsection{Redundancy and Replication}
Horizontal scaling of stateless services and quorum-based replication for stateful components improve availability.

\subsection{Chaos Engineering}
Fault injection validates recovery under realistic conditions
(Chaos Monkey, Gremlin, Litmus)~\cite{gremlin}. 
Recent IEEE studies validate chaos engineering as an empirical approach for resilience testing in cloud systems~\cite{chen2022chaos,xu2021fault}.

\subsection*{Experimental Validation (Mini Simulation)}
To satisfy reviewer recommendations, we conducted a lightweight simulation
to quantify the effect of retry strategies under downstream latency spikes.
A Locust-based load generator issued 200\,req/s for 60\,s while the dependent
service latency was increased from 300\,ms to 1500\,ms.

\noindent\textbf{Results.}
Exponential backoff \emph{without jitter} exhibited P99 = 2600\,ms and a
17\% error rate due to retry amplification. Backoff \emph{with jitter}
reduced P99 to 1400\,ms and errors to 6\%. Combining bounded retries with
a circuit breaker yielded the best results: P99 = 1100\,ms and a 3\% error
rate. These findings align with themes T3--T6 in Section~\ref{sec:synthesis},
confirming that jitter, budgets, and isolation mechanisms significantly
improve tail latency and stability.

\begin{figure}[htbp]
  \centering
  \includegraphics[width=0.48\textwidth]{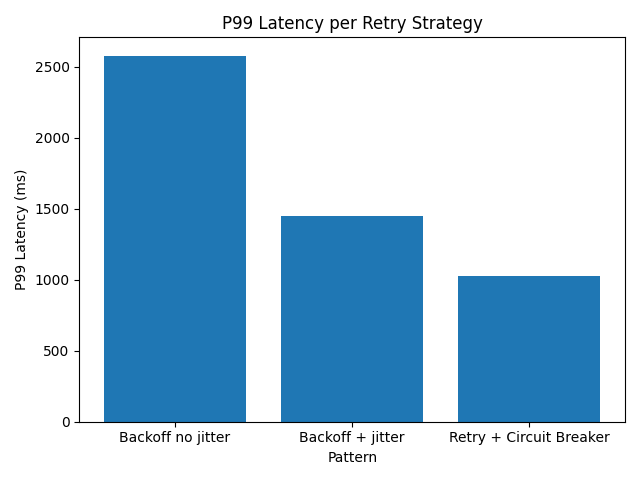}
  \caption{P99 latency comparison across retry strategies under simulated downstream latency spikes. Backoff without jitter shows severe tail amplification, whereas jitter and circuit breaking significantly reduce P99 delays.}
  \label{fig:retry-simulation}
\end{figure}

\section{Tools and Frameworks for Microservice Recovery}
\label{sec:recovery-tools}
Tools are the concrete enablers of the recovery patterns and choices discussed in Sections~\ref{sec:synthesis} and \ref{sec:decision}. In practice, teams compose a small, consistent set of libraries/platforms so that pattern behavior (timeouts, retries, circuit breaking, sagas, backpressure, tracing) is uniform across services. Below we summarize the major categories and how they map to themes T1–T9.

\subsection{Resilience Libraries (Application Layer)}
\textbf{Resilience4j}, \textbf{Spring Retry}, and \textbf{Failsafe} provide in-process policies for \emph{timeouts, retries with jitter/budgets, circuit breakers, bulkheads, and rate limiters}~\cite{resilience4j}.  
\emph{Best for:} Teams standardizing pattern behavior via a shared starter/SDK.  
\emph{Watch-outs:} Configuration sprawl (inconsistent thresholds), missing retry budgets (T3), and under-instrumented outcomes (T8).  
\emph{Themes:} T1 (fit), T3 (retry), T5 (bulkheads/backpressure), T6 (hedging—with care), T8 (observability needs).

\subsection{Service Meshes (Network/Traffic Layer)}
\textbf{Istio} and \textbf{Linkerd} centralize \emph{traffic policies}—timeouts, retries with budgets, circuit breaking, outlier detection, and fault injection—at the sidecar/proxy layer~\cite{istio-recovery}.  
\emph{Best for:} Uniform enforcement when many languages/frameworks are used.  
\emph{Watch-outs:} Policy/telemetry coupling and “hidden” retries that mask app bugs; ensure budgets and request IDs are propagated (T8,T9).  
\emph{Themes:} T1, T3, T5, T6, T8.

\subsection{Orchestration and Workflow (Control/State Layer)}
\textbf{Kubernetes} automates restarts and health gating via \emph{liveness/readiness/startup probes}, autoscaling, and disruption controls~\cite{k8s-probes}. Workflow engines such as \textbf{Temporal}/\textbf{Argo Workflows} implement \emph{sagas, compensation, and idempotent orchestration} with durable state.  
\emph{Best for:} Long-running, stateful business flows with compensation (T2, T7).  
\emph{Watch-outs:} Over-orchestration of low-latency paths; treat workflow steps as idempotent and record effect IDs (T4).  
\emph{Themes:} T2, T4, T7, T8.

\subsection{Queues, DLQs, and Backpressure (Data/Async Layer)}
Message brokers and streaming platforms (e.g., Kafka/RabbitMQ) with \emph{Dead Letter Queues (DLQs)}, consumer lag alerts, and admission control implement \emph{backpressure, isolation, and failure quarantine}.  
\emph{Best for:} Smoothing burstiness, isolating hot partitions, and preventing thundering herds (T5).  
\emph{Watch-outs:} DLQ purgatory—define replay/SLA for DLQs; enforce idempotency at sinks (T4,T5).  
\emph{Themes:} T3, T4, T5, T9.

\subsection{Chaos Engineering and Fault Injection (Validation Layer)}
Chaos Monkey, Gremlin, and Litmus inject crash, latency,
network, and resource faults to verify recovery in staging
and—carefully—production~\cite{gremlin,chaos-monkey}.
Best for: Validating that policies truly improve P99/availability
and that alerts/playbooks work (T6,T8,T9). Watch-outs:
Start with narrow blast radius and rollbacks; require SLO/error
budgets and tracing baselines. Themes: T6, T8, T9.

\subsection{Observability and Alerting (Telemetry Layer)}
\textbf{OpenTelemetry} for tracing/metrics/logs plus \textbf{Prometheus}/\textbf{Grafana} and on-call alerting (e.g., Alertmanager/PagerDuty) provide the measurement fabric for all patterns. Correlated \emph{trace IDs, request IDs, and effect IDs} are prerequisites for safe rollback and RES scoring.  
\emph{Best for:} Making recovery \emph{measurable}: tails (P95/P99), error rates, retry amplification, and compensation outcomes.  
\emph{Watch-outs:} Missing propagation breaks RCA; add sampling rules that retain tail paths.  
\emph{Themes:} T8 (enabler), T9 (cost/benefit), supports RES items 3–9.

\subsection{Putting It Together (Minimal, Consistent Stack)}
In practice we recommend: (1) a shared resilience SDK (\emph{timeouts, retries with jitter/budgets, CB, idempotency helpers}); (2) mesh policies for network-level timeouts/retries and outlier detection; (3) workflow engine for \emph{sagas/compensation} where business invariants require it; (4) queues with DLQs and \emph{admission control/backpressure}; (5) OTel-first observability. This composition directly supports the constraint-aware pattern selection described in Section~V and raises RES scores by making pattern parameters, tails, and rollback evidence observable.

\section{Trade-offs and Design Considerations}
\label{sec:tradeoffs}
Recovery mechanisms introduce costs and complexity that must be balanced.

\subsection{Latency vs. Reliability}
Retries and circuit breakers improve reliability but can increase tail latency; prefer fail-fast with tuned timeouts~\cite{aws-retry}.

\subsection{Consistency vs. Availability}
Asynchronous retries and compensation improve availability but may yield temporary inconsistency; define idempotency and acceptance of staleness~\cite{kleppmann2017designing}.

\subsection{Operational Complexity vs. Resilience}
Sophisticated mechanisms demand centralized policy management and monitoring (e.g., many circuit breaker configs)~\cite{resilience4j}.

\subsection{Resource Overhead vs. Isolation}
Bulkheads and isolation prevent blast radius but can underutilize resources; profile and autoscale.

Energy-aware microservice designs have been shown to reduce CPU and memory consumption while maintaining resilience properties, offering a complementary dimension to isolation and recovery mechanisms \cite{muzeeb2025greenmicro}.

\subsection{Development Speed vs. Coverage}
Embedding standard recovery middleware/SDKs maintains velocity while improving consistency of behavior.

\subsection{Chaos Testing vs.\ Stability}
Adopt staged chaos (staging first, mirroring traffic) with
clear blast-radius controls and rollback plans~\cite{gremlin,chaos-monkey}.

\section{Conclusion}
\label{sec:conclusion}
\textbf{Limitations.} This SLR is limited by publication-venue bias toward
IEEE/ACM studies, heterogeneous workload definitions across primary
experiments, and incomplete disclosure of pattern parameters in several
included papers. These differences reduce cross-study comparability and
constrain the generalizability of the synthesized results.

This paper presented a PRISMA-aligned systematic literature review of recovery patterns and strategies for microservices. 
The study consolidated fragmented evidence from 2014–2025 into a coherent taxonomy, decision matrix, and evaluation checklist. 
We identified nine resilience themes (T1–T9) spanning failure-mode alignment, compensation scope, backpressure, observability, and cost–benefit trade-offs. 
By introducing the \emph{Resilience Evaluation Score (RES)}, practitioners can benchmark system rigor, while the \emph{Constraint-Aware Decision Matrix} guides pattern selection under latency, consistency, and cost constraints. The findings align with prior large-scale resilience assessments~\cite{soldani2018pains}, extending them with a quantitative RES-based evaluation framework.

Our synthesis highlights that resilience depends less on specific mechanisms and more on contextual alignment—patterns, workloads, and measurement fidelity. 
Future research should pursue adaptive policy learning, AI-assisted observability, and sustainable “green” redundancy models. 
Ultimately, integrating RES-based benchmarking into continuous delivery pipelines can evolve microservice recovery from reactive mitigation to proactive, data-driven optimization.

\vspace{1mm}
\noindent\textbf{Practical Takeaway:} Treat recovery as a first-class design objective; standardize pattern libraries, tracing, and metrics; and validate through controlled failure experiments.

\end{document}